\documentclass[superscriptaddress,amsmath,amssymb,floatfix]{revtex4-2}

\usepackage{graphicx}
\usepackage{dcolumn}
\usepackage{bm}
\usepackage{xcolor}
\usepackage{hyperref}
\usepackage{multirow}
\usepackage{tikz}
\usetikzlibrary{quantikz}
\usepackage{mathrsfs}
\usepackage{amsmath}

\newcommand{\ceil}[1]{\lceil #1 \rceil}

\begin{document}


\title{A native measurement-based QAOA algorithm, applied to the MAX $K$-CUT problem}

\author{Massimiliano Proietti}
\altaffiliation{These two authors contributed equally}
\affiliation{%
 Leonardo Labs, Quantum technologies lab, Via Tiburtina, KM 12,400 - Rome - 00131 - Italy
}%
\author{Filippo Cerocchi}%
\altaffiliation{These two authors contributed equally}
 \affiliation{%
 Leonardo Cyber \& Security Solutions division, Via Laurentina 760 - Rome -00143 - Italy
}%
\author{Massimiliano Dispenza}
\affiliation{%
 Leonardo Labs, Quantum technologies lab, Via Tiburtina, KM 12,400 - Rome - 00131 - Italy
}%


\begin{abstract}
Photonic quantum computers, programmed within the framework of the measurement-based quantum computing (MBQC), currently concur with gate-based platforms in the race towards useful quantum advantage, and some algorithms emerged as main candidates to reach this goal in the near term. Yet, the majority of these algorithms are only expressed in the gate-based model of computation, which is incompatible with photonic platforms. Methods to translate gate-based algorithms into the MBQC framework exist, but they are not always optimal in terms of resource cost. In our work, we propose an MBQC algorithm to run the Quantum Approximate Optimization Algorithm (QAOA). Furthermore, we apply the MBQC-QAOA algorithm to the MAX $K$-CUT problem, working for all values of $K$, expressing the cost Hamiltonian and its constraints in a form easily implementable in the MBQC model. We conclude analyzing the resource-cost of our algorithm, compared to the case of translating a gate-based QAOA algorithm into MBQC rules showing up to a 30-fold improvement. With our work, we contribute to close the gap between gate-based and MBQC near-term algorithms, a gap not reflecting the current status of the hardware development.

\end{abstract}

\maketitle

\section{\label{sec:intro}Introduction}
\begin{figure}
\includegraphics[scale=0.9]{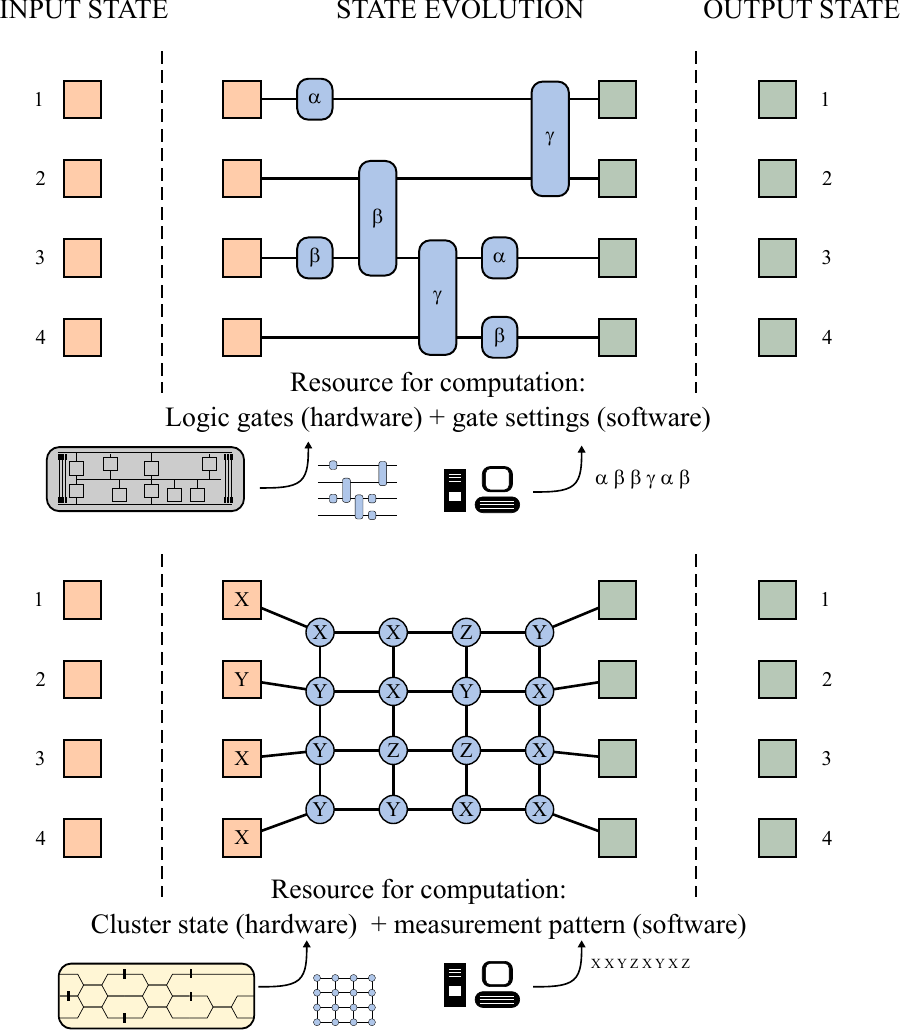}
\caption{\label{fig:scenario}Gate-based (top) and MBQC (bottom) models compared. Both models are universal, and map an input state into an output state of the same dimension. The state evolution instead is fundamentally different for the two models: gate-based models use as a resource for the computation a set of 1-qubit or 2-qubits logic gates (typically provided by a superconductive circuits), and the evolution is driven by the settings for each gate. MBQC models, instead, use as a resource a large cluster state (typically provided by a photonic platform) on which single-qubit measurements are performed according to a pattern given by the user.}
\centering
\end{figure}

MBQC is a framework for universal quantum computation~\cite{raussendorf2001one} developed as an alternative to the gate-based model. Rather than processing the quantum information through a deep sequence of logical gates, it prescribes the use of a large entangled state (the cluster state) on which the computation is executed by sequential adaptive local measurements. The algorithm is encoded on the measurement pattern itself, see Fig~\ref{fig:scenario}. This model of computation is well tailored for, but not limited to~\cite{strydom2022measurement}, photonic platforms~\cite{walther2005experimental,chen2007experimental}. Quantum advantage with a photonic platform was recently demonstrated~\cite{zhong2020quantum, Xanadu_madsen2022quantum} and industrial players are advancing the hardware and the architecture~\cite{arrazola2021quantum,bartolucci2021creation} also giving remote access to their platforms~\cite{Xanadu_madsen2022quantum}. Photonic platforms are limited by photon losses but are highly robust to enviromental noise, they can work at room temperature, they can be fully integrated on chip and can exploit unique multiplexing techniques to generate large cluster states~\cite{yokoyama2013ultra,bombin2021interleaving}. These features, make photonic platforms an ideal candidate for all those in-field applications whereby remote access to the computational source is not granted or allowed, and an offline platform for the quantum computation is required locally, on-board, in a possibly noisy environment. To exploit these platforms, developing new near-term algorithms in the MBQC model is paramount, equivalently to what is the experience for gate-based quantum computation. A class of algorithms drawing a lot of attention recently, are hybrid quantum-classical computational algorithms~\cite{vqa_review_erezo2021variational,vqa_review_moll2018quantum}, which are envisioned to achieve quantum advantage for practical problems in the near future~\cite{qadv_guerreschi2019qaoa,qadv_farhi2016quantum}. Examples of near-term algorithms included in this category are variational quantum eigensolvers (VQEs)~\cite{vqaexp_peruzzo2014variational}, variational quantum simulators (VQSs)~\cite{vqs_benedetti2021hardware} and quantum approximate optimization algorithms (QAOA)~\cite{qaoa_farhi2014QAOA,qaoa_variation_zhu2020adaptive}. In particular, the QAOA was proposed to approximately solve combinatorial optimization problems as the MAX $K$-CUT ~\cite{quoa_app_gaur2008,maxkcut_hadfield2019quantum,maxkcut_wang2020x}, an NP-complete~\cite{goldschmidt1994polynomial} problem which can be used as a primer to approach both fundamental problems in statistical physics~\cite{maxcut_app_barahona1988application} and practical problems such as data clustering~\cite{maxcut_app_poland2006clustering}, and scheduling problems~\cite{maxcut_app_zhang2001budgeted,maxcut_app_yao2007new}. 

Obtaining a formulation of such algorithms in the MBQC model is not trivial. As both the MBQC and the gate-based models are universal, one could think to obtain an MBQC expression by translating known gate-based algorithms. In this case, the mapping between the two universal models was originally introduced in~\cite{raussendorf2001one}, while other mapping methods were later introduced~\cite{danos2007measurement}. However, taking gate-based algorithms as the starting point to obtain MBQC patterns, might not be the optimal strategy. Following the intuition in Ref~\cite{ferguson2021measurement} it is interesting to question whether reasoning \textit{natively} in the MBQC framework, rather than merely translating from circuits could be more efficient. This is the case, for example, for the unitary evolution of diagonal operators in the computational basis, which results to be more efficient if directly implemented in the MBQC model~\cite{browne2016one}, rather than first expressed in the gate-based model and then translated into the MBQC framework.

Here, we find a native MBQC algorithm to solve the MAX $K$-CUT problem via the QAOA. We find the general expression of the cluster state required as a resource for computation, and the pattern of local measurements implementing the target algorithm. To do so, we do not make use of any translation routine from gate-based circuits, but we make use of the efficient formulation within the MBQC of diagonal unitary evolution. In Sec.~\ref{sec:maxkcut}, we first obtain a diagonal expression for the target Hamiltonian encoding the optimization problem. In particular, the form of the Hamiltonian is generalised to any $K$ adding a penalty term, and explicitly expressed only in terms of Pauli $Z$ operators. In Sec.~\ref{sec:mbqc-qaoa} we formulate the general form of the cluster state and measurement pattern, to generate within the MBQC framework the unitary evolution according to the target Hamiltonian, as required by the QAOA algorithm. We apply the method to a simple graph, on which we find a MAX $4$-CUT. Finally, in Sec.~\ref{sec:impl} we evaluate the resource cost of our algorithm, as compared to MBQC algorithms obtained by translating a QAOA gate-based circuit. Remarkably, we show up to a $30$ fold improvement in terms of the cluster state dimension. We conclude in Sec.~\ref{sec:conclusions} by highlighting the next steps towards the application to industrial use-cases as well as pointing out how MBQC algorithms trade-off computational dimensionality with very low depth in terms of computational steps. When accounting for noise, fundamentally limiting depth of gate-based circuits, this feature could be highly relevant in the noisy intermediate-scale quantum (NISQ) era.      

\section{\label{sec:maxkcut} MAX $K$-CUT problem}

The weighted MAX $K$-CUT is a NP-complete problem (see for example \cite{maxkcut_complexity_papadimitriou}) that can be stated as follows: let  $\mathcal G=(\mathcal V, \mathcal E,\underline w_{\mathcal E})$ be an undirected graph with weighted edges where $\mathcal V=\{v_1,...,v_{|\mathcal V|}\}$, $\mathcal E=\{e_1,...,e_{|\mathcal E|}\}$ and let $\mathcal{P}= \mathcal P _1,..., \mathcal P_K$ be a partition  of its vertex set $\mathcal V$ into $K$ classes:
$\mathcal V=\mathcal P_1\cup\cdots \cup\mathcal P_K$.
We shall also refer to such a partition as to a $K$-cut. Let $\mathcal E_{\mathcal{P}}$ be the set of those edges of $\mathcal E$ whose vertices lie into distinct sets of the $K$-cut and let
\small
\begin{equation}\label{Eq: objective_function}
\Delta_K(\mathcal{P})=\sum_{e\in\mathcal E_{\mathcal{P}}} \underline{w}_{\mathcal E}(e).
\end{equation}
\normalsize
The weighted MAX $K$-CUT problem thus consists in finding $\text{argmax}_{\mathcal{P}}(\Delta_K)$ for the function $\Delta_K$ among the $K$-cuts.\\

As mentioned above, finding an optimal solution for the MAX $K$-CUT is a NP-complete problem. As such, there are not efficient classical algorithm to find the optimal solution unless P=NP. For what concerns classical approximation algorithms approaches through semi definite programming (SDP) were proposed to relax the integer programming (IP) problem underlying the MAX $2$-CUT (\cite{maxcut_goesemans}, see also \cite{max_k_cut_improved_algorithm}). However it has been proved \cite{maxkcut_Kann1996OnTH} that for MAX $K$-CUT the relative error between the optimal solution and the solution that can be found by  a classical polynomial time approximation algorithm is lower bounded by $\frac{\delta}{2(K-1)}$ where $\delta$ is the relative error for MAX $2$-CUT. On the other hand it is known \cite{maxcut_inapproximability_Hastad} that approximating MAX $2$-CUT within any constant below $17/16$ is NP-hard, which makes NP-hard approximating MAX $K$-CUT within $1+1/[32(K-1)]-\varepsilon$ for any $\varepsilon>0$. The classical inapproximability of this combinatorial optimization problem makes MAX $K$-CUT a suitable candidate for demonstrating quantum advantage. The approach used to find an  approximate solution to MAX $2$-CUT in \cite{qaoa_farhi2014QAOA}, is to encode the cost function into a suitable target Hamiltonian and try to construct a discretized version of a quantum adiabatic evolution. Interpolating between the (known) maximal energy eigenstate of an initial Hamiltonian and the maximal energy eigenstate of the target Hamiltonian encoding the cost function, the solution of the problem can be found. \\

When it comes to model the instance of the MAX $K$-CUT problem in a way which allows the use of the quantum approximate optimization ansatz \cite{maxkcut_hadfield2019quantum} we need first to choose the encoding. Literature presents three choices: \textit{one-hot} encoding~\cite{oneHot_wang2020x}, \textit{binary} encoding~\cite{maxkcut_fuchs2021efficient} and \textit{qudit} encoding~\cite{qudit_weggemans2022solving}. In the following we focus on the binary encoding.

We assign to every vertex  a label $i$ in $\{0,..,K-1\}$ which can be encoded in states involving $m=\ceil{\log_2(K)}$ qubits, i.e. the smallest integer greater than or equal to $\log_2(K)$. Each class corresponds to the computational basis state given by the binary digits of the index. In particular any $K$-cut $\mathcal{P}$ can be represented as a binary string of length $m|\mathcal V|$ where bits in position $(j-1)m+1$ to $jm$ encodes the class assigned to the $j$th vertex. We shall denote the corresponding $m|\mathcal V|$-state in the computational basis as $\ket{\mathcal{P}}$.

Finding an approximate solution to the MAX $K$-CUT problem requires two ingredients: the \lq\lq mixer", {\it i.e.} a Hamiltonian whose maximal energy state also called the \textit{ansatz state} is well known (and possibly easy to realize), and the target Hamiltonian. Following the original approach of QAOA \cite{qaoa_farhi2014QAOA} we chose the $X$-mixer:
\begin{equation}\label{Eq: X-mixer}
H_m=\sum_{j=1}^{|\mathcal V|}\sum_{\ell=1}^{m} X_{(j-1)m+\ell}
\end{equation}
where we denoted $m=\ceil{\log_2(K)}$ with $X_{(j-1)m+\ell}$ the $X$-Pauli operator on the $\ell$th qubit describing the $j$th vertex (for any choice of indexing on $\mathcal V$) and identity elsewhere. The \textit{ansatz state} of the $X$-mixer on the other hand is well known and it is the following state:
\begin{equation}\label{Eq: ansatz_state}
    \ket{+}^{\otimes m\,|\mathcal V|}
\end{equation}

\begin{figure*}
\includegraphics[scale=0.9]{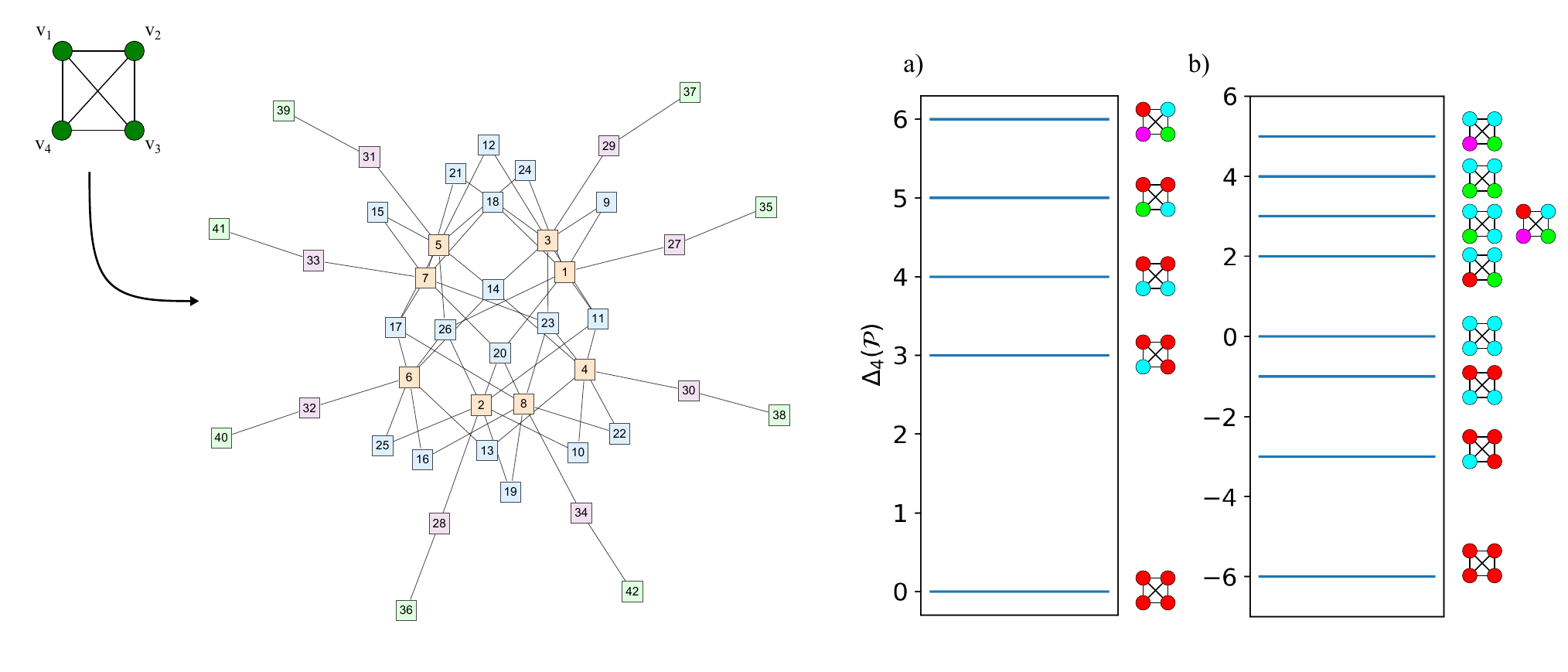}
\caption{\label{fig:cluster state} Max $4$-CUT of a complete graph. The cluster state required for the MBQC-QAOA with $p=1$ is shown on the left. The orange qubits (labeled from $1$ to $8$) are the input and output of $e^{-i\gamma H_t}$. The light blue nodes (labeled from $9$ to $26$) are instead the ancillae $\ket{\phi}_\mathcal{A}$ which have to be measured on the $ZY$-base at angle depending on $\gamma$. The purple qubits (labeled from $27$ to $34$)  extend the computation, mapping the output orange qubits into the green ones (from $35$ to $42$) by the unitary $e^{-i\beta H_m}$. The 8-qubit output state encodes the solution of the MAX $4$-CUT. This is clarified on the right figure where the energy  levels of $H_t$ are shown in a) for the case of $K=4$, while in b) adding the penalty as in Eq~\eqref{Eq:penalisedTarget}. Beside the energy levels we show the corresponding solution. As expected, the top-level energy in a) is the desired solution obtained after classical optimisation of $\gamma$ and $\beta$ and for $p\rightarrow\infty$. In b) we instead demoted all solutions including a red vertex in $\mathcal{P}$, generating new energy levels. The top-energy level is now a solution for the MAX $3$-CUT, while the $K=4$ solution is demoted as shown in the figure. The cluster state including the evolution $H_p$ is not shown.}
\centering
\end{figure*}

We write a Hamiltonian $H_t$ as a weighted sum of smaller terms, one for each edge in the graph:

\begin{equation}\label{Eq: target_Hamiltonian}
H_t=\sum_{\substack{e_k\in\mathcal E}} \underline{w}_{\mathcal E}(e_k) H_{e_k}
\end{equation}
where
\small
\begin{equation}\label{Eq: H_e_Hamiltonian}
H_{e_k}=\frac{2^m-1}{2^m}\cdot{\mathrm{Id}}^{\otimes 2m}
-\frac{1}{2^m}\left(\sum_{\ell=1}^{2^m-1} Z_{v_{i_k}}^{\ell_1,...,\ell_m}\otimes Z_{v_{j_k}}^{\ell_1,...,\ell_m}\right)
\end{equation}
\normalsize
where $\ell=\ell_m\cdots \ell_1$ is the expression of $\ell$ in binary digits, $e_k=\{v_{i_k}, v_{j_k}\}$ and $Z_{v_{i_k}}^{\ell_1,...,\ell_m}=Z_{(v_{i_k}, 1)}^{\ell_1}\otimes\cdots\otimes Z_{(v_{i_k},m)}^{\ell_m}$ proviso that $Z^0=\mathrm{Id}$ (expression for $Z_{v_{j_k}}^{\ell_1,...,\ell_m}$ is analogous). Here by $Z_{(v_{i_k},j)}$ we mean the $Z$-Pauli gate operating on the $j$th qubit describing the vertex $v_{i_k}$. The number of summands is thus equal to $2^m-1$. In the Appendix A, we prove analytically that the terms $H_{e_k}$ are built in such a way that they are the identity on those $2m$-qubit states $\ket{w}\otimes\ket{u}$ such that $\ket{u}\ne\ket{w}$ and annihilate $2m$-qubit states of the form $\ket{w}\otimes\ket{w}$ (here $\ket{u}$ and $\ket{w}$ are $m$-qubit states). The Hamiltonian $H_t$ defined above is precisely the target Hamiltonian in the special case where $K=2^m$.\\
Note that for the case where $K=2^m$, $H_t$ has been chosen in such a way that for any assignment of the vertices to the groups {\it i.e.} for any $K$-cut $\mathcal{P}$, the corresponding $m\,|\mathcal V|$-qubit state  $\ket{\mathcal{P}}$ satisfies
\begin{equation}\label{Eq: H_t_AND_Delta_K}
\Delta_K(\mathcal{P})=\bra{\mathcal{P}}H_t\ket{\mathcal{P}}
\end{equation}
that is the target Hamiltonian has been tailored in order to behave as the function $\Delta_K$ that we want to maximize on $K$-cuts.\\

To address the cases where $K\neq2^m$ we include a penalty term having an analogous expression solely in terms of $Z$-Pauli gates. In Appendix B, we present a recursive approach to the computation of the penalty term which allows for easier manipulation and simplification of the expression. In the following, we will present a non-refined, non-simplified version of the penalty term:
\small
\begin{equation}\label{Eq:penalty}
    H_P=\sum_{e_k\in\mathcal E}\underline{w}_{\mathcal E}(e_k) (H_{v_{i_k}}\otimes\mathrm{Id}+\mathrm{Id}\otimes H_{v_{j_k}}- H_{v_{i_k}}\otimes H_{v_{j_k}})
\end{equation}
\begin{equation}
    H_{v_k}=\frac{2^m-K}{2^m}\mathrm{Id}^{\otimes m}-\frac{1}{2^m}\sum_{j=1}^K \mathcal Z_{v_k, j}
\end{equation}
\normalsize
with
\small
\begin{equation}
    \mathcal Z_{v_k,j}=\sum_{n=1}^{m}\sum_{1\le \ell_n<\cdots <\ell_1\le m} (-1)^{j_{\ell_n}+...+j_{\ell_1}} Z_{v_k, \ell_n}\cdots Z_{v_k,\ell_1}
\end{equation}
\normalsize
where, as usual $j=j_m...j_1$ is the binary digit representation of $j$. The term has been conceived to penalize those quantum states assigning each vertex to a group between $K+1$ and $2^m$: for each edge the penalty term weighs for $\underline{w}_{\mathcal E}(e_k)$ when at least one of the vertices belong to classes from $K+1$ to $2^m$. We can thus define the penalized target Hamiltonian (i.e. the target Hamiltonian for general $K$'s) as:
\begin{equation}\label{Eq:penalisedTarget}
    \widetilde{H}_t=H_t- H_P
\end{equation}
Note that we do not use the same technique used in \cite{maxkcut_fuchs2021efficient}. For example, in \cite{maxkcut_fuchs2021efficient} the authors solve the MAX $3$-CUT problem introducing a pair of ancillae in order to make \lq\lq undistinguishable" assignation to classes $3$ or $4$. The use of those ancilla qubits has a shortcoming: either it increases the circuit depth (this is the choice made by the authors in \cite{maxkcut_fuchs2021efficient}, which add to the total amount of resources only a pair of ancillae) or it increases the number of qubits needed of a multiplicative factor.

\section{\label{sec:mbqc-qaoa} MBQC-QAOA for the MAX $K$-CUT}

The MBQC model requires two ingredients: a cluster state $\ket{\mathcal{C}}$ (typically a large entangled state) and a pattern $\mathcal{M}$ of local measurements on that state. Given a target $n$-qubit unitary evolution $U_{t}$, the challenge is to find which cluster state and which measurement pattern encode that evolution. Additionally, as measurements are probabilistic, a set of local corrections $\mathcal{B}$ depending on the measurement outcomes is required to obtain $U_t$. These are always by-product of the form $Z^i X^j$ with $Z$ and $X$ are the Pauli operators and $i,j$ the values of the measurement outcomes. 

To avoid confusion, in the following when referring to the terms graph, vertices and edges we always mean the graph for which we want to find the MAX $K$-CUT. When instead we refer to nodes and links, we refer to the cluster state of the MBQC.

In our specific scenario, we require two parametric evolution given by $U_t = e^{-i\beta H_m}e^{-i\gamma H_t}$ with $H_t$ from Eq.~\eqref{Eq: target_Hamiltonian} and $H_m$ from Eq.~\eqref{Eq: X-mixer}. The QAOA algorithm states that~\cite{qaoa_farhi2014QAOA} given the parametric unitary evolution
\begin{equation}\label{Eq:QAOA}
    \ket{\vec{\gamma},\vec{\beta}}=\prod_{p=1}^{N_{\text{layer}}} e^{-i\gamma_p H_t} e^{-i\beta_p H_m} \ket{+}^{\otimes n}
\end{equation}
there always exist values for $\gamma_p$ and $\beta_p$ such that the state converges for $N_{\text{layer}}\rightarrow\infty$ to the eigenstate of $H_t$ with maximum eigenvalue as in Eq.~\eqref{Eq: H_t_AND_Delta_K}. However, in practice, finding $\gamma_p$ and $\beta_p$ is not straightforward and is the main challenge for the classical optimization algorithm and is currently an open debate whether this is a strong limitation to achieve quantum advantage~\cite{qaoa_limits_akshay2020reachability}.

In our work the problem we want to solve with the MBQC-QAOA is the MAX $K$-CUT as explained in section~\ref{sec:maxkcut}. We show now how to obtain the cluster state $\ket{\mathcal{C}}$, measurement pattern $\mathcal{M}$ and by-product corrections $\mathcal{B}$. 

For simplicity we set $p=1$ and we start by the parametric evolution $e^{-i\gamma_1 H_t}$. First, we prepare $|\mathcal{V}|m$ qubits in the state $\ket{+}$. We obtain the state $\ket{\psi}_{\mathcal{V}}$ given by
\begin{equation}
    \ket{\psi}_{\mathcal V}=\bigotimes_{\substack{v_j\in\mathcal{V},\\l\in[1,m]}} \ket{+}_{(v_j,l)}
\end{equation}
We then prepare $(K-1)|\mathcal{E}|$ ancillary qubits in the state $\ket{\phi}_{\mathcal A}$ given by
\begin{equation}
    \ket{\phi}_{\mathcal A}=\bigotimes_{\substack{e_k\in\mathcal{E},\\l\in[1,K-1]}} \ket{+}_{(e_k,l)}
\end{equation}
In graphical terms, the total initial state $\ket{\psi}_{\mathcal V}\ket{\phi}_{\mathcal A}$ is represented by $ m|\mathcal{V}|+(K-1)|\mathcal{E}|$ unconnected nodes.

We need now to connect the nodes according to the evolution in Eq~\eqref{Eq: target_Hamiltonian}. Notably, there exists a simple rule~\cite{browne2016one} to find such links if the unitary evolution is diagonal as in Eq~\eqref{Eq: target_Hamiltonian} and also for the penalty in Eq~\eqref{Eq:penalty}. The operator $e^{i \frac{\underline w_{\mathcal E}(e)}{K} Z^{\ell_1,...,\ell_m}\otimes Z^{\ell_1,...,\ell_m}}$ in Eq.\eqref{Eq: H_e_Hamiltonian} has the following form
\begin{equation}\label{Eq:diagonalOp}
    \prod_{j\,:\,\ell_j\ne 0}\exp\left(i\frac{\underline w_{\mathcal E}(e)}{K} Z_{(v_1,j)}Z_{(v_2,j)}\right)
\end{equation}

In particular, any pair of qubits in $\ket{\psi}_{\mathcal V}$ coupled by a $Z$ operator as prescribed in Eq~\eqref{Eq: target_Hamiltonian}, gives rise to a set of $CZ$-gates controlling a common ancilla. We shall explain this in detail in the next paragraph.\\

In our case according to Eq.~\eqref{Eq: target_Hamiltonian} we set $\phi_{k, 1}=\gamma_1 \underline w_{\mathcal E}(e_k)/K$ for every $k=1,..,|\mathcal E|$. In the general case where more layers are required, say $p$, we shall denote the angles $\gamma_j\underline{w}_{\mathcal E}(e_k)$ by $\phi_{k, j}$.
Following the prescription in \cite{browne2016one} we can construct the cluster state as follows: let $\ell=\ell_{m}\cdots\ell_1$ be the representation of $\ell$ in binary digits ($\ell_i$ being the coefficient of $2^{i-1}$). For each $e_k\in \left\{1,..., |\mathcal E|\right\}$ let $e_k=\{v_{i_{k}}, v_{j_k}\}$ and for each $\ell_j\ne 0$, we  apply a $CZ$ operation to the ancillary qubit $\ket{+}_{(e_k,\ell)}$ for each of the corresponding vertex qubits $\ket{+}_{(v_{i_k},j)}$, $\ket{+}_{(v_{j_k},j)}$.

In graphical terms corresponds to connect the ancillary qubits associated to each  summand in Eq. \eqref{Eq: H_e_Hamiltonian} different from the identity to the vertex qubits that are involved in the $Z^{\ell_1,...,\ell_m}\otimes Z^{\ell_1,...,\ell_m}$ operation.  
In general, the $2m(K-1)$-qubit cluster state generated by the edge $e_k$ is 
\begin{equation}
    \mathcal{A}_{e_k} \ket{e_k}= \left(\prod_{\ell=1}^{K-1}\prod_{\ell_i\ne 0}CZ_{(i_k, j),( k, \ell)}CZ_{(j_k, j), (k, \ell)}\right)\ket{e_k}
\end{equation}
where
\begin{equation}\label{Eq: single_edge_initial_state}
    \ket{e_k}=\bigotimes_{j=1}^{m}\bigotimes_{\ell=1}^{K-1}\ket{+}_{(v_{i_k},j)}\ket{+}_{(v_{j_k},j)}\ket{+}_{(e_k,\ell)}
\end{equation}
and where the operator of the form $CZ_{(i_k,j),(k,\ell)}$ perform the control $Z$ operator onto qubits $\ket{+}_{(v_{i_k},j)}\otimes\ket{+}_{(e_k,\ell)}$ leaving unchanged the remaining qubits.
When considering all vertices of the graph we have
\begin{equation}\label{Eq: cluster state}
    \ket{\mathcal{C}_{\gamma_1}}=\prod_{k=1}^{|\mathcal E|}\mathcal{A}_{e_k}\ket{\psi}_{\mathcal V}\otimes\ket{\phi}_{\mathcal A}
\end{equation}
The state in Eq.~\eqref{Eq: cluster state} should now be chiseled into the target $\ket{\gamma_1}=e^{-i\gamma_1H_t}\ket{\psi}_\mathcal{V}$ by locally measuring all the ancillary qubits $\ket{\phi}_{\mathcal A}$ in $\ket{\mathcal{C}_{\gamma_1}}$. These are simultaneously measured in the $YZ$-basis according to the $\phi_{k,1}$ angles defined above. Specifically, for each $\ket{+}_{(e_k,\ell)}$ we measure in the base $\Pi_{\phi_{k,1}}$ given by
\begin{equation*}
    \Pi_{\phi_{k,1}}=\left\{\frac{\ket{+}_{(e_k,\ell)}+e^{i\phi_{k,1}}\ket{-}_{(e_k,\ell)}}{\sqrt{2}},\right.
    \left.\frac{\ket{+}_{(e_k,\ell)}-e^{i\phi_{k,1}}\ket{-}_{(e_k,\ell)}}{\sqrt{2}}\right\}
\end{equation*}
where each element of the base leads to outcome $o[(e_k,\ell)]=\{0,1\}$ respectively. 

Following this measurement round the remaining qubits are in the state $\ket{\gamma_1}$ up to local $Z$-corrections depending on the measurement outcomes given by
\begin{equation}
    \mathcal{B_\gamma}=\prod_{e_k\in|\mathcal{E}|}\prod_{\ell}^{K-1}\prod_{\ell_i\neq0} Z_{(v_{i_k},j)}^{o[(e_k,\ell)]} Z_{(v_{j_k},j)}^{o[(e_k,\ell)]}
\end{equation}

\begin{figure*}
\includegraphics[scale=1]{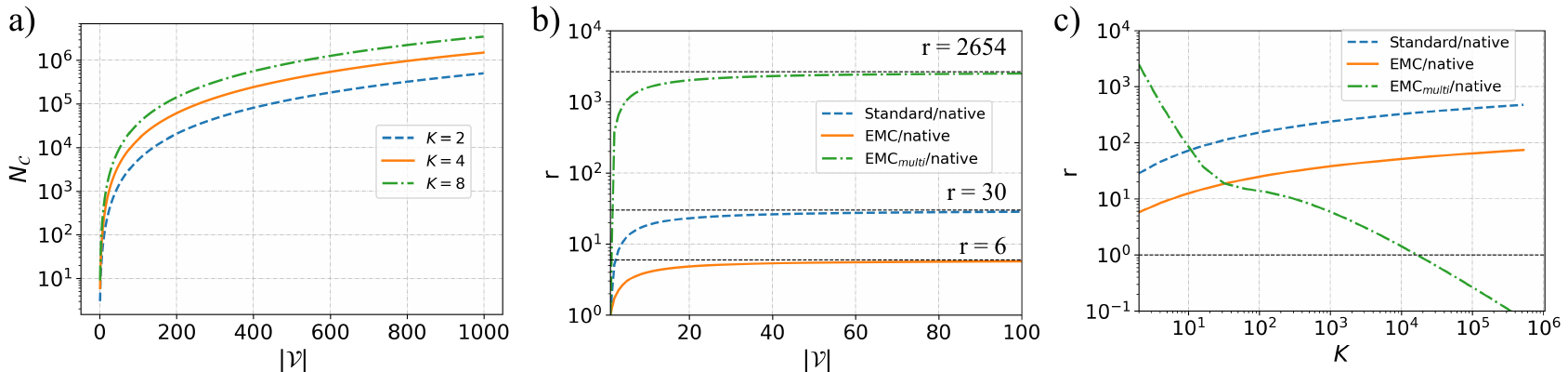}
\caption{\label{fig:results}Resource cost comparison for a complete graph with $|\mathcal{V}|$ vertices. a) We plot the size of the cluster state $N_\mathcal{C}^{\text{native}}$ for different values of $K$ as a function of $|\mathcal{V}|$. In b) and c), instead, we plot the ratio $r$ of the cluster state size required in the standard method (solid orange) and in the EMC method (dashed blue) also applied to the optimised circuit with multi-control gates (dot-dashed green), over the size of the cluster state required for our native approach. In b) we fix $K=2$ and vary $|\mathcal{V}|$, while in b) we fix $|\mathcal{V}|=100$ and vary $K$. Up to $K\approx10^4$ implementing our native approach is more efficient than any strategy. Thereafter, translating the circuit with multi-control gates via the EMC method is a better option, at least in terms of cluster state dimension.}.
\centering
\end{figure*}

The remaining evolution given by $e^{-i\beta_1 H_m}$ is straightforward as it only contains local terms. In particular it requires each qubit in $\ket{\gamma_1}$ to be entangled with two additional qubits in a chain-like cluster state according to
\begin{equation}\label{Eq:beta_state}
    \bigotimes_{\substack{v_j\in\mathcal{V},\\l\in[1,m]}} CZ_{a_1,a_2}CZ_{(v_j,l),a_1}\ket{\gamma_1}_{(v_j,l)}\ket{+}^{a_1}_{(v_j,l)}\ket{+}^{a_2}_{(v_j,l)}
\end{equation}
the qubit $(v_j,l)$ is measured in the $X$-basis with outcome $o[(v_j,l)]$, and the ancilla $a_1$ generated by qubit $(v_j,l)$ (which we label in Eq.~\eqref{Eq:beta_state} as $\ket{\,}^{a_1}_{(v_j,l)}$) is subsequently measured in the $XY$-basis at angle $\beta_1(-1)^{o[(v_j,l)]}$. The qubit $a_2$ is the output qubit, and it is corrected with by-product $X^{o[(a_2)]}Z^{o[(v_j,l)]}$. We have shown the evolution of $e^{-i\beta_1 H_m}e^{-i\gamma_1 H_t}$ as two separate steps, but applying the rules from Ref~\cite{danos2007measurement} it is possible to move the by-product corrections $\mathcal{B}_\gamma$ after the entangling operations in Eq~\eqref{Eq:beta_state}, obtaining a global cluster state where the input state is $\ket{\psi}_\mathcal{V}$ and the output state $\ket{\gamma_1,\beta_1}$. In this case, we require only three computational rounds to generate the target evolution i.e. a simultaneous measurement of all qubits in $\ket{\mathcal{C}_{\gamma_1}}$, and a subsequent simultaneous measurement of all qubits $a_1$ with an angle depending on outcomes of the previous round, concluding with a serial of local $XZ$ corrections. Remarkably, the number of rounds does not depend on the graph dimensionality. Increasing the number of vertices or edges in the graph we want to cut, increases the number of ancillae, but as the measurements can be done simultaneously the computational steps are unchanged. This is in contrast with the gate-based model, where the addition of edges or nodes requires new gates in turn increasing the circuit depth.      

To appreciate the algorithm just described, we apply the protocol to an unweighted four vertices complete graph as shown in the top left of Fig~\ref{fig:cluster state}. We consider $K=4$ giving $m=2$ in the binary encoding. We construct the cluster state $\ket{\mathcal{C}_{\gamma_1}}$ where the qubits in Fig~\ref{fig:cluster state} labeled from $0$ to $8$ refer to $\ket{\psi_{\mathcal{V}}}$ while the ones labeled from $9$ to $26$ are in fact the ancillae $\ket{\phi_{\mathcal{A}}}$. The cluster state is completed by adding the qubits encoding the evolution according to the $X$-mixer Hamiltonian $H_m$. We obtain a $42$-qubit cluster state as shown in Fig~\ref{fig:cluster state}. 

To demonstrate that the Hamiltonian encoding chosen is correct, in Fig~\ref{fig:cluster state} a) we show the energy levels of $H_t$ for the graph we are considering. As shown in the figure, the top energy level corresponds to the state
\begin{equation*}
    \ket{0}_{(v_1,1)}\ket{0}_{(v_1,2)}\otimes\ket{0}_{(v_2,1)}\ket{1}_{(v_2,2)}\otimes
    \otimes\ket{1}_{(v_3,1)}\ket{0}_{(v_3,2)}\otimes\ket{1}_{(v_4,1)}\ket{1}_{(v_4,2)}
\end{equation*}
which in fact identifies the expected solution. Note that all the permutations of such state with respect to $v_i$ are degenerate. For completeness, we also set $K=3$ and $m=2$ and add the penalty term introduced in Eq~\eqref{Eq:penalty}. The energy levels of the new Hamiltonian $\widetilde{H_t}$ are shown for comparison in Fig~\ref{fig:cluster state} b). The top energy state is now the state
\begin{equation*}
    \ket{0}_{(v_1,1)}\ket{0}_{(v_1,2)}\otimes\ket{0}_{(v_2,1)}\ket{1}_{(v_2,2)}\otimes
    \otimes\ket{1}_{(v_3,1)}\ket{0}_{(v_3,2)}\otimes\ket{1}_{(v_4,1)}\ket{0}_{(v_4,2)}
\end{equation*}
corresponding to the solution of the MAX $3$-CUT. In this case the solution found for $K=4$ was demoted into a lower energy level, as shown in the figure. We remark that this approach differs from the strategy in Ref~\cite{maxkcut_fuchs2021efficient}, where the unwanted $K+1$ to $2^m$ classes are made indistinguishable via the use of additional ancillae and at the cost of circuit depth.

\section{\label{sec:impl} Results}

In the following we compare our method with other two different MBQC algorithms to implement the evolution $U_t=e^{-i\beta H_m}e^{-i\gamma H_t}$, without including the penalty Hamiltonian and assuming $K=2^m$, meaning that we only consider $K=\{2,4,8,\dots,2^m\}$ for integer values of $m$. These two algorithms are obtained by translating into a MBQC pattern a gate-based circuit. We consider the circuit shown in Appendix C, expressed in terms of CNOT and $R_Z$ gates only. This, for $K=2^m$, implements $U_t$ when $H_t$ is composed by a sequence of $Z$ operators, as in Eq~\eqref{Eq: target_Hamiltonian}. The first MBQC algorithm is derived by applying the rules in Ref~\cite{raussendorf2001one}, we shall refer to this approach as \textit{standard}. The second is again obtained by translating $U_t$, but by applying rules in Ref~\cite{danos2007measurement}, we shall refer to this as \textit{EMC} due to its structure in terms of entanglement (E), measurement (M) and correction (C) commands. The main difference is that in the standard approach, the CNOT gate is implemented with a $15$-qubit cluster where input and output qubits do not match, while in the EMC case, the CNOT gate is implemented with a $4$-qubit cluster state where one input overlaps with one output. We compare these two with our algorithm, presented in Sec.~\ref{sec:mbqc-qaoa}, which we shall refer as \textit{native}. In our case, there is no translation of logic gates, but we rather implement directly on the cluster state a diagonal evolution. The comparison is performed with respect to the number of qubits required for the creation of the cluster state, which is the resource of the computation. We recall that in our native implementation the cluster state as explained in Sec~\ref{sec:mbqc-qaoa} is composed by $N_\mathcal{C}^{\text{native}}$ qubits given by
\begin{equation}
    N_\mathcal{C}^{\text{native}} = 3|\mathcal{V}|\log_2K+|\mathcal{E}|(K-1)
\end{equation}
In Fig.~\ref{fig:results} a), we show $N_\mathcal{C}^{\text{native}}$ for different values of $K$ and varying $|\mathcal{V}|$.

To compute the resource cost of both the standard and EMC case, we first count the number of CNOT and $R_z$ gates for the circuit in Appendix C, these are given by 
\begin{equation}\label{Eq: CX_gates}
\#\text{CNOT}=|\mathcal E|\cdot\sum_{\ell=1}^{m}\binom{m}{\ell}2 (2\ell-1)=|\mathcal E| [2(\log_2K-1)K+2]
\end{equation}
\begin{equation}\label{Eq: R_Z_gates}
\# R_Z=|\mathcal E|\cdot\sum_{\ell=1}^{m}\binom{m}{\ell}=|\mathcal E| (K-1)
\end{equation}
Moreover, considering also $e^{-i\beta H_m}$ we need to add a layer of $R_X$ gates for each qubit
\begin{equation}\label{Eq: R_X_gates}
\# R_X=|\mathcal V|\log_2K
\end{equation}
We now introduce $N_{C}$, $N_Z$ and $N_X$ as the number of qubits required by an MBQC pattern to implement a CNOT gate, a $R_Z$ gate and a $R_X$ gate respectively. In particular, $N_{C}=15$ and $N_{Z}=5$ for the standard approach, while $N_{C}=4$ and $N_{X}=3$ for the EMC approach. For an MBQC algorithm on $N_i=m|\mathcal{V}|$ input qubits, the cluster state size is given by
\begin{equation}
    N_\mathcal{C}^{\text{Gate}\rightarrow\text{MBQC}} = N_i + \#\text{CNOT}\cdot(N_{C}-2) +\#R_Z\cdot(N_{Z}-1) + \#RX\cdot(N_X -1)
\end{equation}
which should be compared to $N_\mathcal{C}^{\text{native}}$.
In Fig.~\ref{fig:results} b) and c) we show the ratio $r=N_\mathcal{C}^{\text{Gate}\rightarrow\text{MBQC}}/N_\mathcal{C}^{\text{native}}$ for both the standard (blue dashed curve) and EMC (orange solid curve) methods. We do this by considering a complete graph, for which it holds $|\mathcal{E}|=|\mathcal{V}|(|\mathcal{V}|-1)/2$, in Fig.~\ref{fig:results} b) by fixing $K=2$ while varying $\mathcal{V}$ and in Fig.~\ref{fig:results} c) by fixing $|\mathcal{V}|=100$ while varying $K$. The more general asymptotic expressions are given in the Appendix D. Notably, as shown in both Fig.~\ref{fig:results} b) and c) our native method shows a large improvement in terms of resource cost. It requires up to a factor $30$ less qubits for the generation of the cluster state when compared to the standard approach, and a factor $5$ compared to the EMC method. For increasing $K$ this advantage can be even greater as shown in c). This is obtained for complete graphs, we note that in applications of the MAX $K$-CUT for data clustering a complete graph is required, making our result relevant for practical applications of the QAOA algorithm. 

We note that the circuit we considered for the comparison, has a number of CNOT and $R_Z$ gates growing linearly with $K$ (exponentially with the binary encoding qubits $m$). Although in practical applications, as for example MAX $K$-CUT for clustering, $K$ is typically on the order of $10$, one might ask if a more efficient circuit to translate exists. It is the case in Ref~\cite{maxkcut_fuchs2021efficient} where the same target Hamiltonian $H_t$ is decomposed in terms of few CNOT gates per edge, assisted by one multi-control gate leading to a total number of CNOT gates scaling logarithmically with $K$~\cite{multi_controlled_liu2008} (polynomial on $m$), see Appendix C for more details. 
For completeness we also apply the EMC rules to translate such $K$-optimised circuit, and compare with our native MBQC algorithm. This is shown with a green dot-dashed curve in Fig.~\ref{fig:results} b) and c). It is interesting to see that, for $K<16384$ the native method is still more efficient, then turning to be more convenient translating the circuit in \cite{maxkcut_fuchs2021efficient} via the EMC rules. This is a consequence of the polynomial dependence obtained in this case. 

We note that, our analysis as in Fig.~\ref{fig:results}, does not account for the measurement rounds required by the MBQC algorithm. We stress that, in our native algorithm, the evolution $e^{-i\gamma H_t}$ can be implemented with only two measurement rounds, independently on $|\mathcal{V}|$, $|\mathcal{E}|$ and $K$. When translating a circuit into a MBQC pattern, the circuit depth inherently influences the measurement rounds of the obtained MBQC algorithm. A quantitative study regarding computational depth complexity is beyond the scope of this work, but it is where we envision the greatest advantage of our method. We leave this to future work.  

We compared the resource cost only considering the target Hamiltionian $H_t$ without penalty. We note however, that in Ref~\cite{maxkcut_fuchs2021efficient} to address the cases with $K\neq2^m$, they introduce additional ancillae and non-diagonal 2-qubit gates. As explained Sec~\ref{sec:maxkcut} we instead provide a diagonal form of $H_p$, and therefore we expect an even larger improvement in this case. The computation of the general number of gates required as in Eq~\eqref{Eq: CX_gates} is more complex and was not computed in this work as it does not change the behavior of $r$.

\section{\label{sec:conclusions} Conclusions}

In our work, prompted by the fast development of photonic platforms requiring algorithmic reasoning in the MBQC model, we showed a new measurement-based protocol to implement the QAOA algorithm applied on the generalised MAX $K$-CUT. We showed this in detail, generalising as much as possible all the expressions obtained and compared its resource cost to other MBQC algorithms. The main contribution is the expression of the QAOA algorithm natively in the MBQC model, where diagonal operators have a nice and highly efficient implementation. To do this, we provided an expression of both the target and the penalty Hamiltonian only in terms of Pauli $Z$ operators. This approach, for $K=2$, leads to a $30$ fold improvement with respect to standard MBQC methods and a $6$ fold improvement with respect to optimized MBQC methods. For larger values of $K$, up to the order of $K\approx10^4$ our method is more convenient even if compared with a circuit optimised for large values of $K$. For $K>10^4$ using such optimised circuit embedded with optimised MBQC mapping rules can be a more efficient strategy, although only in terms of resource size.  

In practical use cases, where for example the MAX $K$-CUT is exploited for data clustering, $K$ is typically on the order of $10$ and our native method outperforms other strategies. This is highly relevant in the NISQ era, where only a limited amount of resources are available, and methods to optimize their use are key to achieve useful results. For this reason, we expect our approach to stimulate theoretical and experimental investigations of native MBQC approaches.

On this line, we recall that one major roadblock for useful quantum computation is the computational depth. This is related to the circuit depth in the gate-based model, and to the measurement rounds in the MBQC framework. Regarding the comparison with other MBQC algorithms, we expect a significant improvement as in our case the number of measurements is reduced to the minimum. A more detailed and quantitative comparison is left to future work. More generally, it is certainly interesting to also compare MBQC and gate-based models. Intuitively, we can see the former as algorithms requiring high computational cost in terms of space (in the form of the cluster state dimension) while manifesting very few computational rounds (in the form of measurements depending on previous outcomes). The latter, instead, seems to behave in the opposite direction. In practice, noise imposes constraints on time and space, and it is therefore interesting to compare the two approaches on practical use cases and scenarios.

\begin{acknowledgments}
We wish to acknowledge Carlo Liorni for useful discussions on the proposed method, and Sebastiano Corli for useful discussions on the MBQC algorithm. 
\end{acknowledgments}

\appendix

\section{Proof for target Hamiltonian}

Let $\mathcal G=(V,E)$ be a graph. We fix $m\in\mathbb N$ and let $K=2^m$ and we label each vertex in $V$ with numbers from $0$ to $K-1$ represented in bits. We then think at those numbers as vectors from the computational basis of $\mathcal H_m=(\mathbb C^{\otimes 2})^{\otimes m}$. Let us consider now the following function $\mathcal O_m:\mathcal H_m\otimes \mathcal H_m\rightarrow \mathcal H_m$:
\begin{align*}
\mathcal O_m=&\frac{K-1}{K}\cdot\left({\mathrm{Id}}^{\otimes m}\otimes {\mathrm{Id}}^{\otimes m}\right)-\\&\frac{1}{K}\left(\sum_{\substack{w\in\{a,b\}^*, |w|=m,\\ w\ne a\cdots a, \varepsilon}} w(\mathrm{Id}, Z)\otimes w(\mathrm{Id}, Z)\right)
\end{align*}
where we denoted by $\{a,b\}^*$ the set of finite words in the alphabet $\{a,b\}$, by $|\cdot|$ the word length in the alphabet $\{a,b\}$ and by $\varepsilon$ the empty word in $\{a,b\}^*$.\\

{\bf Theorem.} {\it For any $m\in\mathbb N$ we have that:
\begin{align*}
&\mathcal O_m( |i_1\cdots i_m\rangle\otimes |j_1 \cdots j_m\rangle )=\\&=\left\{
\begin{array}{c}
|i_1\cdots i_m\rangle\otimes |j_1\cdots j_m\rangle \mbox{ {\rm if} } |i_1\cdots i_m\rangle\ne |j_1\cdots j_m\rangle\\
\mathbf 0 \mbox{ {\rm otherwise} }
\end{array}
\right.
\end{align*}}
\vspace{3mm}

{\bf Proof.} We proceed by induction on $m$. For $m=1$ the operator $\mathcal O_1$ is equal to:
$$\mathcal O_1=\frac{\mathrm{Id}\otimes\mathrm{Id}}{2}-\frac{Z\otimes Z}{2}$$ 
A straightforward computation shows that the matrix representing $\mathcal O_1$ over the computational basis $\{|00\rangle, |01\rangle, |10\rangle, |11\rangle\}$ of $\mathcal H_1\otimes\mathcal H_1$ is:
$$
\left(
\begin{array}{cccc}
0 & 0 & 0 & 0\\
0 & 1 & 0 & 0 \\
0 & 0 & 1 & 0\\
0 & 0 & 0 &0 
\end{array}
\right)
$$
We now assume the statement true for $m-1$. We prove that it holds true for $m$ as well. We first write $\mathcal O_m$ in a different way:
\small
\begin{align*}
&\mathcal O_m= \frac{2^m-1}{2^m}\cdot\left(({\mathrm{Id}}^{\otimes (m-1)}\otimes{\mathrm{Id}})\otimes({\mathrm{Id}^{\otimes(m-1)}}\otimes\mathrm{Id})\right)-\\&-\frac{1}{2^{m}}\cdot\left(\sum_{\substack{w\in\{a,b\}^*,\\ |w|=m-1,\\ w\ne\varepsilon}} (w(\mathrm{Id}, Z)\otimes{\mathrm{Id}})^{\otimes 2}+(w(\mathrm{Id}, Z)\otimes Z)^{\otimes 2}\right)-\\
& -\frac{1}{2^m}\cdot(\mathrm{Id}^{\otimes(m-1)}\otimes Z)^{\otimes 2}
\end{align*}
\normalsize
In order to prove that the statement holds we shall consider several cases. Let us consider first the case where the argument is $|i_1\cdots i_m\rangle\otimes |i_1\cdots i_m\rangle$ we look at the terms under the sum sign: in this case, every summand is equal to $2\cdot |i_1\cdots i_m\rangle^{\otimes 2}$ so that the terms inside the square brackets becomes:
$$[2(2^{m-1}-1) +1]\cdot |i_1\cdots i_m\rangle^{\otimes 2}=(2^m-1)|i_1\cdots i_m\rangle^{\otimes 2}$$
and we conclude that:
$$\mathcal O_m(|i_1\cdots i_m\rangle^{\otimes 2})=\mathbf 0$$
We shall now consider the case where $\mathcal O_m$ is applied to vectors of the form $v\otimes w$ with $v\ne w$. It will convenient to consider several subcases. To start with, we shall first consider the case where $v=|i_1\cdots i_{m-1}0\rangle$ and $w=|i_1\cdots i_{m-1} 1\rangle$ that is the case where the two sequence differ only for the last bit. In these circumstances we see that:
\begin{align*}
&(w(\mathrm{Id}, Z)\otimes{\mathrm{Id}})^{\otimes 2} |i_1\cdots i_{m-1}0\rangle\otimes |i_1\cdots i_{m-1}1\rangle+\\
+&(w(\mathrm{Id}, Z)\otimes Z)^{\otimes 2}|i_1\cdots i_{m-1}0\rangle\otimes |i_1\cdots i_{m-1}1\rangle=\mathbf 0
\end{align*}
Hence:
\begin{align*}
&\mathcal O_m(|i_1\cdots i_{m-1}0\rangle\otimes|i_1\cdots i_{m-1}1\rangle)=\\&=\frac{K-1}{K}\cdot |i_1\cdots i_{m-1}0\rangle\otimes |i_1\cdots i_{m-1}1\rangle-\\&-\frac{1}{K} \left(-|i_1\cdots i_{m-1}0\rangle\otimes |i_1\cdots i_{m-1}1\rangle\right)=\\&=|i_1\cdots i_{m-1}0\rangle\otimes |i_1\cdots i_{m-1}1\rangle
\end{align*}

An analogous argument allows to draw the same conclusion for the evaluation of $\mathcal O_m$ over vectors of the form $|i_1\cdots i_{m-1}1\rangle\otimes |i_1\cdots i_{m-1}0\rangle$.\\ Let us now consider the case $|i_1\cdots i_{m-1}i_m\rangle\otimes|j_1\cdots j_{m-1}i_m\rangle$ with $|i_1\cdots i_{m-1}\rangle\ne |j_1\cdots j_{m-1}\rangle$. In this case each summand under the sum sign is equal to
$$2((w(\mathrm{Id},Z)|i_1\cdots i_{m-1}\rangle)\otimes|i_m\rangle)\otimes((w(\mathrm{Id},Z)|j_1\cdots j_{m-1}\rangle)\otimes|i_m\rangle)$$
now let us consider the isomorphism 
\small
$$T: (\mathcal H_{m-1}\otimes \mathcal H_1)\otimes(\mathcal H_{m-1}\otimes \mathcal H_1)\rightarrow (\mathcal H_{m-1}\otimes\mathcal H_{m-1})\otimes (\mathcal H_1\otimes \mathcal H_1)$$
$$T((\mathbf v_i\otimes \mathbf w_j)\otimes (\mathbf v_k\otimes \mathbf w_\ell))=(\mathbf v_i\otimes \mathbf v_k)\otimes (\mathbf w_j\otimes \mathbf w_\ell)$$
\normalsize
Let us define the following operators:
\small
$$W_{\mathrm{Id},Z}=\sum_{\substack{w\in\{a,b\}^*,\\ |w|=m-1, w\ne\varepsilon}} (w(\mathrm{Id},Z)\otimes\mathrm{Id})^{\otimes 2}+(w(\mathrm{Id},Z)\otimes Z)^{\otimes 2}$$
$$
V_{\mathrm{Id},Z}=\sum_{\substack{w\in\{a,b\}^*,\\ |w|=m-1, w\ne\varepsilon}} w(\mathrm{Id},Z)^{\otimes 2}\otimes(\mathrm{Id}^{\otimes 2}+ Z^{\otimes 2})
$$
$$
V_{\mathrm{Id},\mathrm{Id}}=\sum_{\substack{w\in\{a,b\}^*,\\ |w|=m-1, w\ne\varepsilon}} w(\mathrm{Id},Z)^{\otimes 2}\otimes(\mathrm{Id}^{\otimes 2}+ \mathrm{Id}^{\otimes 2})
$$
\normalsize
Then we have the following chain of equalities:
\small
$$-\frac{1}{2^{m}}\cdot W_{\mathrm{Id},Z}\left(|i_1\cdots i_{m-1}i_m\rangle\otimes |j_1\cdots j_{m-1}i_m\rangle\right)=$$
$$
=T^{-1}\circ\left(-\frac{1}{2^m}\cdot V_{\mathrm{Id},Z}\right)\circ T\left(|i_1\cdots i_{m-1}i_m\rangle\otimes |j_1\cdots j_{m-1}i_m\rangle\right)=
$$
$$= T^{-1}\circ\left(-\frac{1}{2^m} V_{\mathrm{Id},\mathrm{Id}}\right)\circ T\left(|i_1\cdots i_{m-1}i_m\rangle\otimes |j_1\cdots j_{m-1}i_m\rangle\right)=$$
$$
=T^{-1}\circ\left(\mathcal O_{m-1}-\frac{2^{m-1}-1}{2^{m-1}}\cdot \mathrm{Id}^{\otimes2(m-1)}\right)\otimes\mathrm{Id}^{\otimes 2}\circ$$
$$\circ\, T\left(|i_1\cdots i_{m-1}i_m\rangle\otimes |j_1\cdots j_{m-1}i_m\rangle\right)=
$$
$$=\frac{1}{2^{m-1}}|i_1\cdots i_{m-1}i_m\rangle\otimes|j_1\cdots j_{m-1}i_m\rangle$$
\normalsize
Where the second equality holds in restriction to vectors of the prescribed form (i.e. vectors whose last bit coincide) while the third equality holds by the inductive hypothesis. Summing all terms we get:
\small$$\left[\frac{2^m-1}{2^m}+\left(\frac{1}{2^{m-1}}-\frac{1}{2^m}\right)\right]\cdot|i_1\cdots i_{m-1}i_m\rangle\otimes |j_1\cdots j_{m-1}i_m\rangle$$
\normalsize
which gives us the correct result. 
Finally we consider the subcase where the last qubit is different: $v=|i_1\cdots i_{m-1}0\rangle$, $w=|j_1\cdots j_{m-1}1\rangle$. We immediately observe that the summands under the sum sign cancel out and the terms left give the desired result. $\Box$

\section{Recursive approach for the computation of penalty Hamiltonian }

Let us consider a $1$-qubit system $\mathcal H=\mathrm{Span}_{\mathbb C}(\ket{0},\ket{1})$. We can write the projection onto the subspaces $\mathbb C\ket{0}$ and $\mathbb C\ket{1}$ respectively as $\ket{0}\bra{0}, \ket{1}\bra{1}$ using the bra-ket formalism. Moreover, it is easy to write these projections in terms of the identity matrix and of $Z$-Pauli gates:

$$\ket{0}\bra{0}=\dfrac{(\mathrm{Id}+Z)}{2};\quad \ket{1}\bra{1}=\dfrac{(\mathrm{Id}-Z)}{2}$$

Using these expression we are able to compute the projection on specific vectors of the computational basis of a $m$-qubit system $\mathcal H^{\otimes m}$ in terms of the $Z$-Pauli gate:
\small
\begin{align*}
&\ket{\delta_1\cdots \delta_m}\bra{\delta_1\cdots\delta_m}=\bigotimes_{i=1}^{m}\frac{(\mathrm{Id}+(-1)^{\delta_i}Z)}{2}=\\&=\frac{1}{2^m}\bigotimes_{i=1}^{m}(\mathrm{Id}+(-1)^{\delta_i}Z)=\\&=\frac{1}{2^m}\left(\mathrm{Id}^{\otimes m}+\sum_{i=1}^{m}\sum_{1\le\ell_1<\cdots<\ell_i\le m} (-1)^{\delta_{\ell_1}+\cdots +\delta_{\ell_i}} Z_{\ell_1}\cdots Z_{\ell_i}\right)
\end{align*}
\normalsize
where we use the notation $Z_{\ell_1}\cdots Z_{\ell_i}$ to denote the operator acting as follows on vectors $v_1\otimes \cdots \otimes v_m$:
\small
$$(Z_{\ell_1}\cdots Z_{\ell_i})(v_1\otimes\cdots\otimes v_{\ell_1}\otimes\cdots\otimes v_{j}\otimes\cdots\otimes  v_{\ell_i}\otimes \cdots \otimes v_k )=$$
$$=v_1\otimes\cdots\otimes Z_{\ell_1}v_{\ell_1}\otimes\cdots \otimes v_j\otimes \cdots\otimes Z_{\ell_i}v_{\ell_i}\otimes \cdots \otimes v_k$$
\normalsize
In order to construct a penalization hamiltonian to force the evolution of a quantum state to be constrained to a given subspace we should be able to find more efficient expressions for projection onto subspaces of $\mathcal H^{\otimes m}$. To this end we shall focus on projections onto subspaces generated by the first $M$ elements of the computational basis and by the last $N-M$ elements with $N=2^m$ (the two cases being equivalent since, if $P$ is the projection onto the first subspace $\mathrm{Id}^{\otimes m}- P$ is the projection onto the second subspace).\\

{\bf Example.} Let us consider the case where $m=2$ and $M=3$. We want to determine the projection onto the subspace generated by the computational basis states $\ket{00}, \ket{10}, \ket{01}$. We want to find an expression for:
$$\ket{00}\bra{00}+\ket{10}\bra{10}+\ket{01}\bra{01}=\mathrm{Id}-\ket{11}\bra{11}$$
Using the expression we found for $\ket{\delta_1\cdots\delta_m}$ we can compute the expression on both sides of the equation:
\small
$$
\ket{00}\bra{00}+\ket{10}\bra{10}+\ket{01}\bra{01}=\frac{\mathrm{Id^{\otimes 2}}}{4}+$$$$+\frac{1}{4}\left(Z_1+Z_2+Z_1Z_2-Z_1+Z_2-Z_1Z_2+Z_1-Z_2-Z_1Z_2\right)=$$$$
=\frac{3}{4}\mathrm{Id^{\otimes 2}}+\frac{1}{4}(Z_1+Z_2-Z_1Z_2)
$$
and
$$
\mathrm{Id}^{\otimes 2}-\ket{11}\bra{11}=\mathrm{Id}^{\otimes 2}-\frac{\mathrm{Id}^{\otimes 2}}{4}-\frac{1}{4}(Z_1Z_2-Z_1-Z_2)
$$
\normalsize
\vspace{3mm}

{\bf Recursive Procedure to compute projections.} Keeping in mind the Example we shall now present a general strategy for computing projections onto subspaces spanned by the firsts $2^{m-1}<M<2^m$ basis states of the computational basis. In order to do so we must first find a way to group the first $M$ computational basis states in a way which is suitable to operate simplifications.\\

By the assumption on $M$ we immediately see that the projections on those state whose $m$th qubit is $\ket{0}$ sum to a projection operator of the following form:
$$\frac{1}{2}(\mathrm{Id}^{\otimes m}+\mathrm{Id}^{\otimes(m-1)}\otimes Z)$$
Let us now consider those states  which are among the $(2^{m-1}+1)$th and the $M$th. These states all have the last qubit set to $\ket{1}$.  By construction this will be a subspace of the $2^{m-1}$-dimensional subspace generated by the computational basis states $\ket{\eta_1\cdots\eta_{m-1}1}$, as $\eta_1,...,\eta_{m-1}\in\{0,1\}$. Let $\ket{\delta_1\cdots \delta_{m-1} 1}$ be the top computational basis state in the projection subspace i.e. the $M$th element of the computational basis. Calling $P_{m-1}$ the projection onto this subspace we immediately see that we get the following expression for the projection:
$$\frac{P_{m-1}\otimes\mathrm{Id}}{2}-\frac{P_{m-1}\otimes Z}{2}$$
We are thus reduced to consider the projection problem on a lower dimensional state space $\mathcal H^{\otimes(m-1)}$, in particular we want to compute the projection onto the subspace generated by the first $0<M-2^{m-1}<2^{m-1}$ elements of the computational basis states. Now if $M-2^{m-1}=2^{\ell}$ for some $0\le\ell<m-1$ we deduce that the projection $P_{m-1}$ can be written as:
$$\mathrm{Id}^{\otimes\ell}\otimes\left(\frac{\mathrm{Id}+Z}{2}\right)^{\otimes(m-1-\ell)}$$
which concludes the construction of the projector onto the $M$-dimensional subspace of $\mathcal H^{\otimes m}$ spanned by the first $M$ eigenvectors of the computational basis.
Otherwise let $m_1$ be the integer such that $2^{m_1-1}<M-2^{m-1}<2^{m_1}$. In this case we apply the same reasoning as above and we split the projection into two parts:
$$P_{m-1}=\mathrm{Id}^{\otimes (m_1-1)}\otimes\left(\frac{1+Z}{2}\right)^{\otimes (m-m_1)}+$$$$+ P_{m_1-1}\otimes\left(\frac{\mathrm{Id}-Z}{2}\right)\otimes\left(\frac{\mathrm{Id}+Z}{2}\right)^{\otimes(m-1-m_1)}$$
where $P_{m_1-1}$ represents the projection on the first $M-2^{m-1}-2^{m_1-1}$ vectors of the computational basis of $\mathcal H^{\otimes m_1-1}$
and we repeat the process for $P_{m_1}$. The following procedure ends in $m$ steps at most. Substituting backwards the expression found for $P_{m-1}$ and $P_{m_i-1}$ for $i=1,..,\ell$ we get final expression for the projection on the desired subspace.\\

{\bf Example.} We shall now test the procedure on the following example. We consider the $7$ dimensional subspace in $\mathcal H^{\otimes3}$ spanned by the first $7$ basis states of the computational basis. The goal is to \lq\lq run" the procedure described above to derive the expression for the projection in terms of the $Z$-Pauli gates. Notice that this computation has an obviously simpler solution which is provided by:
\small
$$\mathrm{Id}^{\otimes 3}-\ket{111}\bra{111}=$$$$=\mathrm{Id}^{\otimes 3}-\frac{1}{8}(\mathrm{Id^{\otimes 3}}-Z_1-Z_2-Z_3+Z_1Z_2+Z_2Z_3+Z_1Z_3-Z_1Z_2Z_3)=$$
$$=\frac{7}{8}\mathrm{Id}^{\otimes 3}+\frac{1}{8}(Z_1+Z_2+Z_3-Z_1Z_2-Z_2Z_3-Z_1Z_3+Z_1Z_2Z_3)$$
\normalsize
We start with the first step:
$P=\frac{1}{2}\left(\mathrm{Id}^{\otimes 3}+\mathrm{Id}^{\otimes2}\otimes Z\right)+\frac{1}{2}\left(P_2\otimes\mathrm{Id}-P_2\otimes Z\right)$.
Now we consider $M=7-4=3$. Following our routine we see that $2<M<4$ and thus we set $k_1=2=k-1$. Using our expression for $P_2$ we get:
$$P_2=\mathrm{Id}\otimes\left(\frac{\mathrm{Id}+Z}{2}\right)+P_1\otimes\left(\frac{\mathrm{Id}-Z}{2}\right)$$
Now we set $M=7-4-2=1$ and we see that $P_2$ is simply the projection on the computational basis state $\ket{0}$ in $\mathcal H$. We substitute backwards:
$$P_2=\mathrm{Id}\otimes\left(\frac{\mathrm{Id}+Z}{2}\right)+\left(\frac{\mathrm{Id}+Z}{2}\right)\otimes\left(\frac{\mathrm{Id}-Z}{2}\right)=$$
$$=\frac{3}{4}\mathrm{Id}^{\otimes 2}+\frac{1}{4}(Z_1+Z_2-Z_1Z_2)$$
The final substitution thus yields:
$$
P=\frac{1}{2}\mathrm{Id}^{\otimes 3}+\frac{1}{2}Z_3+\frac{3}{8}\mathrm{Id}^{\otimes 3}+\frac{1}{8}(Z_1+Z_2-Z_1Z_2)-\frac{3}{8} Z_3-$$$$-\frac{1}{8}(Z_1Z_3+Z_2Z_3-Z_1Z_2Z_3)=
$$
$$
=\frac{7}{8}\mathrm{Id}^{\otimes 3}+\frac{1}{8}(Z_1+Z_2+Z_3-Z_1Z_2-Z_2Z_3-Z_1Z_3+Z_1Z_2Z_3)
$$

\vspace{2mm}

\section{Circuit realizing the unitary $e^{-i\gamma H_t}$}
\begin{figure}[h]
    \centering
    \includegraphics[scale=0.6]{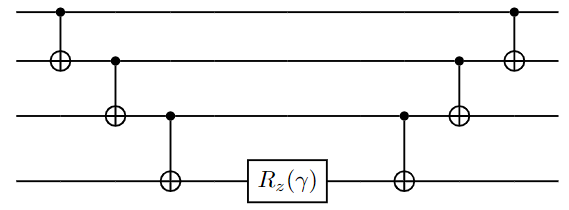}
    \caption{Circuit block for $\exp(-i\gamma ZZZZ)$}
    \label{fig:appendix1}
\end{figure}

According to the expression for the hamiltonian $H_{e_k}$ in the main text, we have to implement terms like Exp$\left[i\gamma\underbrace{ZZ\dots ZZ}_{|\mathcal{V}|m}\right]$. These can be realised by a ladder of $(|\mathcal{V}|m-1)$ CNOT gates, followed by a rotation along the Z-axis with angle $\gamma$ and again by a ladder of $(|\mathcal{V}|m-1)$ CNOT gates. For example, a 4-qubit term like $\exp\left[i\gamma\underbrace{ZZZZ}_{|\mathcal{V}|m}\right]$ can be realised by the circuit in Fig\ref{fig:appendix1}.
The general circuit implementing $e^{-i\gamma H_t}$ follows by adding in sequence all the circuital blocks implementing each of the $K-1$ terms in $H_{e_k}$. 

\begin{figure}[h]
    \centering
    \includegraphics[scale=0.6]{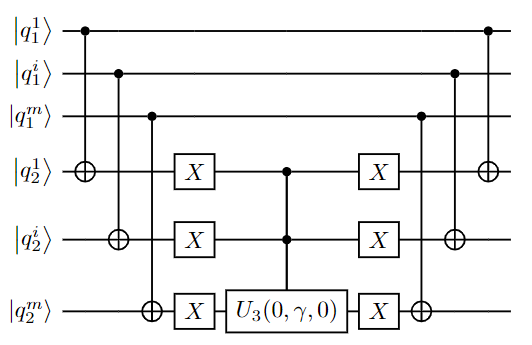}
    \caption{Circuit expression for $\exp(-i\gamma H_{e_k})$ using multi-control gates.}
    \label{fig:appendix2}
\end{figure}

This circuit can be further optimised, to obtain an expression having a number of CNOT gates scaling polynomially with $m$. It is immediate to verify that the circuit in Fig\ref{fig:appendix2} realizes the unitary operator $\exp(-i\gamma H_{e_k})$, up to a global phase $e^{-i\gamma}$. \\

where we are using the notation:
$$
U_3(\varphi,\vartheta, \lambda)=
\left(
\begin{array}{cc}
\cos(\varphi/2) & -e^{i\lambda}\sin(\varphi/2)\\
e^{i\vartheta}\sin(\varphi/2) & e^{i(\vartheta+\lambda)}\cos(\varphi/2)
\end{array}
\right)
$$
The $m$-controlled unitary gate here, can be expressed using the square root $V$ of the operator $U_3(0,\gamma, 0)$ and its adjoint, together with a number of $CX$ gates which is polynomial in $m$. Notably, this approach significantly improve the number of $CX$ gates, but the circuit is no longer parallelizable as the one presented above.

This can be decomposed further into $|\mathcal{E}|(24m^2-210m+540)$ CNOT gates and $|\mathcal{E}|(32m^2-286m+739)$ single-qubit gates resulting polynomial in $m$.

\section{\label{sec:rem} Asymptotic expressions for $r$}

We show in the following the general expression for $r=N_\mathcal{C}^{\text{Gate}\rightarrow\text{MBQC}}/N_\mathcal{C}^{\text{native}}$ for the cases considered in the main text. We recall that in the following $m=\log_2 K$. For the standard approach we have
\begin{equation*}
    r = \frac{|\mathcal{E}| (K (26 m-22)+22)+3 m |\mathcal{V}|}{|\mathcal{E}| (K-1)+3 m |\mathcal{V}|}
\end{equation*}
while for the EMC approach we have 
\begin{equation*}
  r = \frac{|\mathcal{E}| (K (4 m-2)+2)+3 m |\mathcal{V}|}{|\mathcal{E}| (K-1)+3 m |\mathcal{V}|}  
\end{equation*}

If we instead consider the EMC approach applied to the $m$-optimised circuit with multi-control gates we have
\begin{equation*}
    r = \frac{2 |\mathcal{E}| (m (88 m-207)+1446)+3 m |\mathcal{V}|}{|\mathcal{E}|
   (K-1)+3 m |\mathcal{V}|}
\end{equation*}
To obtain a more compact expression we consider comple graphs for which $|\mathcal{E}|=|\mathcal{V}|(|\mathcal{V}|-1)/2$ and then we consider the limit $V\rightarrow\infty$ obtaining for the same three cases respectively
\begin{align*}
    &r=\frac{26 K m-22 K+22}{K-1}\quad(\text{Standard})\\
    &r=\frac{4 K m-2K+2}{K-1}\quad(\text{EMC})\\
    &r=\frac{176 m^2-414 m+2892}{K-1}\quad\quad(\text{EMC m-optimised})
\end{align*}


%

\end{document}